\documentclass{article}
\usepackage{graphicx} 
\usepackage[round, sort&compress, numbers]{natbib}
\usepackage{hyperref}
\hypersetup{
    colorlinks=true,
    linkcolor=blue,
    filecolor=magenta,      
    urlcolor=cyan,
    pdftitle={Overleaf Example},
    pdfpagemode=FullScreen,
    }
\usepackage[margin=1in]{geometry}
\title{The Rise of Bluesky}

\author{Ozgur Can Seckin\textsuperscript{*\rm 1}, Filipi Nascimento Silva\textsuperscript{1}, Bao Tran Truong\textsuperscript{1}, Sangyeon Kim\textsuperscript{1}, 
\\ 
Fan Huang\textsuperscript{2}, Nick Liu\textsuperscript{1}, Alessandro Flammini\textsuperscript{1}, Filippo Menczer\textsuperscript{1} 
\\
{\footnotesize \textsuperscript{\rm 1}Observatory on Social Media, Indiana University, Bloomington, IN, USA}
\\[-0.3em]
{\footnotesize \textsuperscript{\rm 2}Luddy School of Informatics, Computing, and Engineering, Indiana University, Bloomington, USA}
}

\date{}

\begin{document}

\maketitle
\def\thefootnote{*}\footnotetext{To whom correspondence should be addressed: oseckin@iu.edu}

\begin{abstract}
This study investigates the rapid growth and evolving network structure of Bluesky from August 2023 to February 2025. 
Through multiple waves of user migrations, the platform has reached a stable, persistently active user base. 
The growth process has given rise to a dense follower network with clustering and hub features that favor viral information diffusion. 
These developments highlight engagement and structural similarities between Bluesky and established platforms.
\end{abstract}

\section*{Introduction}

Bluesky is a microblogging platform similar to Twitter (now X) in functionality but with decentralized governance and data control. It was launched as an invitation-only service in February 2023 and opened to the general public in February 2024 \citep{forbes2024invite}.
Bluesky experienced rapid growth through multiple bursts of new users, reaching 30 million accounts at the end of January 2025.

The utility of a social media platform increases with user adoption \cite{easley2010network}. 
Such a network effect hinges on the ability to sustain a stable user base. 
This is challenging as most users sign up, browse briefly, and never return \cite{Guardian2023mastodon}.
The open availability of Bluesky data provides an unprecedented lens into the early rise of a major social media platform. We examine how its activity and network structure have co-evolved between August 2023 and February 2025 to investigate whether the platform has reached a stable, self-sustaining state.

\begin{figure}
\centering
\includegraphics[width=\textwidth]{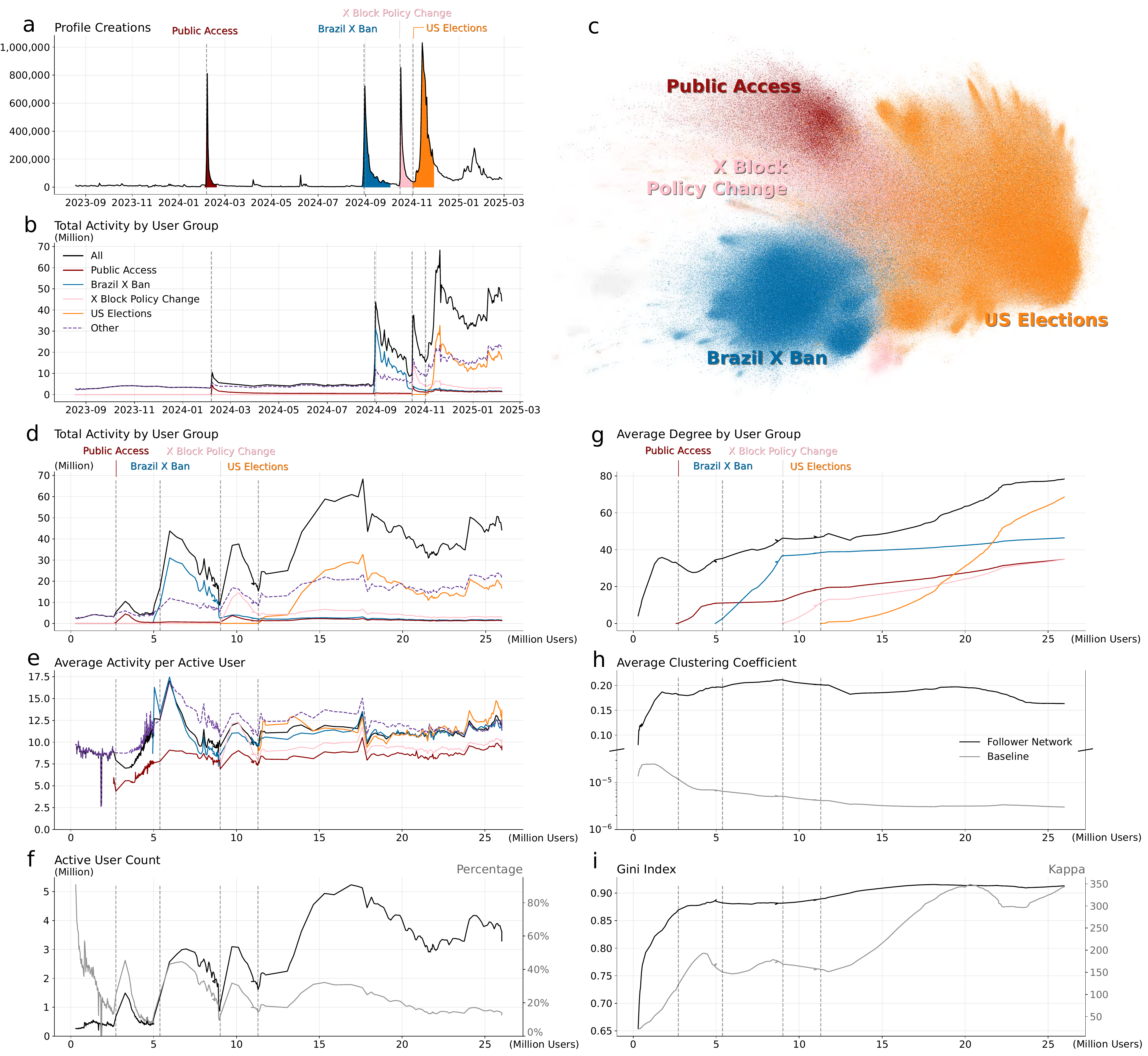}
\caption{Daily engagement and network metrics on Bluesky. Vertical dashed lines represent four major events. (a)~Profile creation occurred in bursts corresponding to these events. We assign users to four groups based on the periods in which they joined (shown by the colors). (b)~Total activity by user group. The dashed line corresponds to users outside of the four major groups. (c)~Visualization of the follower network based on a 1\% sample of the users. Panels (d-i) plot various quantities versus the increasing number of users; only 26 million users with at least one friend or follower as of 7 February 2025 are considered. Note that the number of users can occasionally decrease due to deletion and suspension of accounts. (d)~Total activity by user group. (e)~Average activity per active user, defined as having at least one post, repost, reply, or like. (f)~Daily active users. We plot the absolute count (black) and the percentage out of total users (gray). (g)~Average out-degree by user group.  (h)~Average clustering coefficient of the Bluesky follower network (black). We also plot the expected value in a random network with the same number of nodes and edges, which is given by the average degree divided by the node count (gray). (i)~Gini (black) and Kappa (gray) indices of in-degree heterogeneity.}
\label{mainfig}
\end{figure}

The rise of Bluesky was driven by four major events in 2024, leading to the migration of users from other platforms, as illustrated in Fig.~\ref{mainfig}a. On February 6, Bluesky switched from an invitation-based service to public access. We observe a sharp increase in profile creations around this date, with a majority of new accounts using Japanese.
X was blocked in Brazil on August 30 \cite{guardian2024brazil, verge2024brazil}. Shortly after the ban, we observe nearly half a million accounts created on Bluesky, the great majority using Portuguese and likely Brazilian users migrating from X \cite{verge2024brazil}.
On October 16, X announced a policy change allowing users to view posts from those who had blocked them \cite{forbes2024xchange}, driving a wave of user departures from the platform \cite{verge2024bluesky, cbs2024bluesky}. New profiles on Bluesky during this period predominantly use English.
The platform experienced a more sustained influx of new users after the US Presidential Elections on November 4. The migration of prominent influencers and news sources from X was widely reported \cite{guardian2024}.

\section*{Results}
Given the bursts of migration, we identify four major groups of users based on the periods when they joined Bluesky, as illustrated by the colors in Fig.~\ref{mainfig}a.
Each user influx triggers a surge in activity (posts, reposts, replies, and likes), followed by a decay (Fig.~\ref{mainfig}b). The black line reflects the aggregated activity of these four major groups plus others who joined outside of these waves.

Fig.~\ref{mainfig}c illustrates the follower network based on a small sample of nodes on January 20, 2025. We observe three broad communities, which can be partially attributed to the languages of the user groups: Japanese, Portuguese, and English. 

Given the bursty dynamics of the network, let us analyze the evolution of the platform using ``network time'' (the number of users in the network) rather than the actual timeline. This is equivalent to rescaling time so that it is easier to focus on the periods of rapid growth. In Fig.~\ref{mainfig}d, the two groups driven by platform-related events (Brazil X ban and X block policy change) exhibit similar engagement patterns characterized by a sudden burst followed by a decay. Following the reinstatement of X in Brazil on October 8 \cite{bbc2024banlift}, most of the users in the Brazil X ban group likely left Bluesky and returned to X. The arrival of new groups appears to reinvigorate existing users. For example, following the X block policy change in October, there is a sharp increase in activity across all users.
The users who joined following the US Elections display a different pattern characterized by slower growth and decay, suggesting more sustained activity.

The average activity of users active on the platform is driven by the bursts of incoming users, as shown in Fig.~\ref{mainfig}e. Users in all groups became more active after the X block policy change. The average activity seems to have stabilized above ten actions per day.

The number of daily active users reached a steady state after the US Elections, with a proportion around 15\% of all existing accounts (Fig.~\ref{mainfig}f). This suggests around 85\% users are silent, also known as lurkers, consistent with studies on Twitter \cite{gong2015lurkers, Antelmi2019lurkers, mcclain2021behaviors}.

These activity patterns suggest an overall increase in engagement accumulated through multiple waves. 
To investigate the relationship between activity and connectivity among Bluesky users, let us inspect how the follower network evolves over time.
Since we reconstruct the network based on follow actions, we exclude singleton nodes, i.e., users with neither friends nor followers.
Fig.~\ref{mainfig}g shows that the average out-degree --- mean number of accounts followed per user --- increases over time for all user groups.
Users who joined after the public access did not follow many other accounts, resulting in a decrease in the average degree. 
But in general, users from each new wave increase the average degree, suggesting an ease of integration into the network over time.
Furthermore, new bursts of incoming users boost the following activity of previous waves.
For the most recent users who joined after the US Elections, we observe an acceleration in the number of new connections.
We also analyzed the connectivity of the network (not shown). A giant component consisting of around 70--80\% of all users emerged even before public access. Its relative size has remained fairly stable, suggesting that the network has been effectively integrating new users. 

Fig.~\ref{mainfig}h plots the average clustering coefficient of the undirected version of the follower network, a measure that indicates the presence of tightly connected communities in which two users with a shared connection are likely to be connected to each other. The clustering coefficient is orders of magnitude higher than that of an equivalent random network. This shows that, as is typical in social networks, Bluesky has a strong friend-of-a-friend structure, which facilitates the viral spread of information \cite{weng2013virality}.

Finally, let us examine whether some hub accounts amass a large portion of followers. One way to measure the heterogeneity of the in-degree distribution is through the Gini index, calculated by comparing the cumulative proportions of the nodes to the cumulative proportions of the followers they have. It takes values between zero and one, with larger values indicating higher inequality. Another measure of heterogeneity is the Kappa index, defined as $\kappa = \langle k^2 \rangle / \langle k \rangle^2$, where $k$ is the in-degree and $\langle \rangle$ indicates average. Values greater than one indicate large fluctuations in the degree distribution due to the presence of hubs. Fig.~\ref{mainfig}i plots both Gini and Kappa indices, demonstrating the emergence of highly influential accounts with many followers, consistent with measurements on Weibo and Twitter \cite{han2016comparative}. Both indices display a strong growth in the post-election surge. 

\section*{Discussion}

Our analysis reveals that after multiple waves of user growth, Bluesky has reached a robust active user base with a cohesive follower network and sustained platform engagement. The users who joined following the US Elections exhibit particularly high activity levels and network density, which have persisted beyond an initial spike. We are also observing the emergence of influential accounts. These indicators suggest that Bluesky is becoming structurally similar to established social media platforms. As it achieves mainstream adoption, the platform is likely to attract bad actors and harmful content. This highlights the need for research on potential abuse and manipulation, as well as effective strategies to mitigate these risks.

\paragraph{Data Availability} 

The code and datasets to reproduce the results are available at \url{https://github.com/osome-iu/rise_of_bluesky} and \url{https://zenodo.org/records/15066073}.

\paragraph{Competing Interest Statement}

The authors declare no competing interests.

\paragraph{Acknowledgements:} We are grateful to Ben Serrette and Caitlin Watkins for support. This work was supported in part by the Swiss National Science Foundation (Sinergia grant CRSII5\_209250) and by the Knight Foundation.
We used the IU JetStream 2 computational infrastructure through allocation CIS230183 from the Advanced Cyberinfrastructure Coordination Ecosystem: Services \& Support (ACCESS) program, which is supported by NSF grants 2138259, 2138286, 2138307, 2137603, and 2138296.

\bibliographystyle{unsrt}
\bibliography{bib.bib}
\end{document}